\documentclass{aa}

\usepackage{graphicx}
\usepackage{hyperref}
\usepackage{adjustbox}
\newcommand{\cifig}[1]{Figure~\ref{#1}}

\usepackage{txfonts}
\usepackage{lipsum}

\usepackage{lscape}             
                                
\usepackage{placeins}

\begin{document}
\let\linenumbers\relax
   \title{Dust torques for realistic dust size distributions}

   \author{Vincenzo Roatti\inst{1,2}\fnmsep\thanks{Corresponding author: vincenzo.roatti@phd.unipd.it}
        \and Giovanni Picogna\inst{2}
        \and Francesco Marzari\inst{3}
        }

   \institute{Dipartimento di Fisica e Astronomia "G. Galilei", Università degli Studi di Padova, Vicolo dell'Osservatorio 3, 35122 Padova, Italy
   \and Universitäts-Sternwarte, Ludwig-Maximilians-Universität München, Scheinerstr. 1, München, 81679, Bayern, Germany
   \and Dipartimento di Fisica e Astronomia "G. Galilei", Università degli Studi di Padova, Via Marzolo 8, 35121 Padova, Italy}

   \date{\today}

  \abstract
   {Previous studies have shown that a population of dust particles with a fixed Stokes number can exert a substantial torque on a low-mass planet embedded in a protoplanetary disk, modifying its migration rate.}
   {We aim to characterize the dust torque on a low-mass planet for a realistic distribution of dust grain sizes.} 
   {We performed 2D hydrodynamical simulations of planet-disk interactions using the PLUTO code, with the addition of Lagrangian superparticles representing dust dynamics.  {We apply an energy-based criterion to exclude the particles that are gravitationally bound to the planet, to prevent circumplanetary flow to contaminate the torque measurements.}}
    {We find that the dust torque is dominated by the largest grains in the size distribution and is highly sensitive to the maximum grain size. For typical disk conditions, the torque becomes positive for marginally coupled particles ($\mathrm{St} \gtrsim 10^{-2}$) and can exceed the gas torque in the presence of cm-sized pebbles, leading to outward migration of low-mass planets.  {Unlike previous studies, the turbulent dust diffusion has a negligible influence on the torque over the explored range of $\alpha  = 10^{-4}$ to $3\times 10^{-3}$.} The dominant contribution arises from within the planetary Hill sphere, highlighting the need for high spatial resolution and accurate integration of particle trajectories. We derive a scaling law for the dust torque as a function of the maximum grain size and the planetary mass, suitable for implementation in population synthesis models.}
   {}


   \maketitle

\section{Introduction}

The formation and orbital architecture of planetary systems are fundamentally shaped by the interaction between growing planets and their natal protoplanetary disks. For low-mass planets (terrestrial to super-Earth masses), the exchange of angular momentum with the surrounding gas leads to Type I migration \citep[see the review by][]{2012ARA&A..50..211K}. In standard isothermal models, the combined action of the Lindblad and corotation torques typically results in rapid inward migration with a timescale shorter than the disk lifetime, in contrast with the observed abundance of short-period exoplanets \citep{2023A&A...670A.139R}.

Although most migration studies have focused on the gas component, protoplanetary disks are intrinsically dusty environments. The dust component can exert a dynamical influence on the embedded planets, introducing an additional contribution to the total torque. Using 2D multi-fluid simulations, \cite{benitez2018} showed that a planet can scatter nearby dust particles, generating an asymmetric dust distribution that produces a net torque. For marginally coupled particles (St $\gtrsim 0.1$), a dust-depleted region forms behind the planet, leading to a positive contribution that can partially or fully counteract the negative gas torque, depending on the local dust-to-gas ratio \citep{2018MNRAS.478.2737C, 2020MNRAS.497.2425H}. Similar conclusions were reached by \cite{2024ApJ...972..152H} using linear calculations, emphasizing the importance of accurately resolving the dust dynamics in the vicinity of the planet.

Subsequent studies have explored additional physical effects. Dust accretion onto the planet can enhance positive torques over a broader region of parameter space, both in fluid models \citep{regaly2020} and in simulations employing Lagrangian superparticles \citep{chrenko24}. In contrast, turbulent diffusion tends to smooth the dust distribution, reducing the front–rear asymmetry responsible for the torque, and thus weakening or suppressing its contribution \citep{2025A&A...698A..21C}. 

All the previous works make the major assumption that dust can be described by a single and constant Stokes number. In the Epstein drag regime, this is  {locally} equivalent to assuming that all dust particles have the same size,  {i.e. in a given region where the gas surface density is fixed}. However, dust in protoplanetary disks is expected to follow a continuous size distribution, shaped by coagulation and fragmentation processes, ranging from micrometer-sized grains to centimeter-sized pebbles \citep{2024ARA&A..62..157B}. Since aerodynamic coupling strongly depends on particle size, different grain populations can contribute differently to the total dust torque. Moreover, for marginally coupled particles, the drag regime may vary, and the Stokes number becomes sensitive to local gas properties and relative velocities between the gas and dust component. 

In this work, we compute the dust torque acting on a low-mass planet for a realistic grain size distribution. We perform high-resolution 2D hydrodynamical simulations and model the dust component using Lagrangian superparticles sampling a continuous size distribution. This approach allows us to resolve particle trajectories in the vicinity of the planet, while including a simplified treatment of dust accretion.

Rather than prescribing a constant Stokes number, we fix the physical size of each particle and compute the aerodynamic drag self-consistently from the local gas conditions. This enables us to characterize the dust torque as a function of the maximum grain size in the disk and to assess its impact on planetary migration for a given choice of disk parameters. This paper is organized as follows. In Section \ref{sec:methods} we describe the numerical setup and torque computation. In Section \ref{sec:results} we present and discuss our results. In Section \ref{sec:conclusions} we present our conclusions.

\section{Methods} \label{sec:methods}
We simulated the evolution of a planet embedded in a gas disk using the PLUTO code \citep{mignone2007}, with the addition of Lagrangian particles representing the dust dynamics and an N-body module for the planet (see \cite{2025A&A...703A.270R} for more details). We considered a thin, locally isothermal disk with no self gravity, in which the gas surface density and the disk temperature are power laws in radius:
\begin{align}
    \Sigma _g  & = \Sigma _0 \left( \frac{r}{1 \, \rm{au}} \right) ^{-p} \\
    T & = T_0 \left( \frac{r}{1 \, \rm{au}} \right) ^{-q}.
\end{align}
We investigate three different values for $\Sigma_0 = $ 10, 100 and 1000 g/cm$^2$, assuming fixed p = 0.5. The absolute value of $\Sigma _g$ affects our results by modifying the coupling between the gas fluid and the dust particles of fixed size. For the temperature, we assume $T_0 = 630$ K and q = 1.0, which corresponds to a fixed disk aspect ratio of h = 0.05. For disk viscosity, we use the $\alpha$-prescription \citep{shakura1973} with $\alpha = 10^{-4}$, which is consistent with previous theoretical work \citep{regaly2020, chrenko24} and recent observational constraints \citep{2026MNRAS.548ag423L}. We also consider higher values of $\alpha = 10^{-3}, 3 \times 10^{-3}$ to check the effect of high turbulence on the resulting dust torques \citep{2025ApJ...979..185H, 2025A&A...698A..21C}. We adopted a uniform 2D polar grid extending from 0.7 to 1.3 au in radius and from 0 to 2$\pi$ in azimuth, with the planet located at 1 au in a fixed orbit. The resolution of the grid is chosen to have approximately 10 squared cells per Hill radius of the planet, $r_{H} = (M_{\rm{pl}}/3M_*)^{1/3}$, which corresponds to a 351x3675 grid for a 5 M$_{\oplus}$ planet and a 702x7350 grid for $M_{\rm{pl}}$ = 1 M$_{\oplus}$. We have checked that doubling the resolution to 20 cells per Hill radius does not significantly affect the results (see Appendix~\ref{app:res_conv}). We added wave-killing zones at the inner and outer radial boundaries of the grid, so that the surface density of the gas is damped towards its initial value \citep{2006MNRAS.370..529D}.

For the dust component, we initialize $N = 1.6 \times 10^{7}$ particles between $r_{\rm{min}} = 0.8$ and $r_{\rm{max}} = 1.2$ au, separated into 16 bins of different sizes, ranging from 3.125 $\mu$m to 10.24 cm. These are initially distributed to  {follow} the same  {radial slope as the gas disk}. Both the gas and the dust components evolve at the same time in our code. In particular, 
dust particle orbits are numerically integrated with a semi-implicit scheme \citep{2014ApJ...785..122Z} under the gravitational influence of the planet and the star, while also experiencing local gas drag, which is computed by interpolating the Epstein and Stokes regimes as in \citet{picogna2018}:
\begin{equation} \label{Cd}
    C_D = \frac{9\mathrm{Kn}^2 C _D ^{\mathrm{Eps}} +C _D ^{\mathrm{Stk}} }{(3\mathrm{Kn} + 1)^2}\,,
\end{equation}
where $C _D ^{\mathrm{Eps}}$ and $C _D ^{\mathrm{Stk}}$ are the Epstein and Stokes drag coefficients, respectively, and Kn = $\lambda / 2s$ is the Knudsen number, with $\lambda$ the mean free path of the gas and $s$ the particle size \citep{2003A&A...399..297W, 2009A&A...493.1125L, picogna2018}. From equation \ref{Cd} we can compute the dust stopping time,
\begin{equation} \label{taus}
    \tau _s = \frac{4 \lambda \rho _d}{3 \rho _g C_D c_s}\frac{1}{\mathrm{MaKn}}\,,
\end{equation}
where $\rho _d$ = 3 g/cm$^3$ is the internal density of dust particles,  {$\rho _g = \Sigma _g/(2\pi H)$ is midplane gas density in the thin disk approximation}, $c_s$ is the speed of sound, and $\rm{Ma} = |{v}_{\rm{rel}}|/c_s $ is the Mach number. Then, the drag force acting on a dust grain moving with velocity $\mathbf{v}_{\mathrm{rel}}$ relative to the gas is given by
\begin{equation}
    F_D = - \frac{\mathbf{v}_{\mathrm{rel}}}{\tau _s} .
\end{equation}
When the stopping time is less than ten times the integration timestep, we switch to the fully-implicit integrator \citep{2014ApJ...785..122Z}.
The particles are discarded from the simulation when they cross $r_{\rm{min}}$, and new particles are continuously initialized to maintain a constant dust flux between [0.92, 1]$r_{\rm{max}}$, as in \cite{chrenko24}. Dust feedback onto the gas is neglected, so that the dust mass does not affect the gas dynamics: therefore, we assign a mass to each particle a posteriori, assuming a standard MRN (Mathis-Rumpl-Nordsieck) distribution \citep{MRN}, $n (a) \propto a^{-\beta}$, and a dust-to-gas ratio of $Z = 0.01$,
\begin{equation}
    m_{\rm{part}}(a) = \frac{Z M_{\rm{gas}}}{N(a)}\frac{a^{4-\beta}}{\sum _i a_{i}^{4-\beta}},
    \label{mass}
\end{equation}
where $M_{\rm{gas}}$ is the mass of the gas disk between $r_{\rm{min}}$ and $r_{\rm{max}}$, $N(a)$ is the number of particles of size $a$, $\beta$ = 3.5 and the sum is made over the size bins. The last equation shows that the overall mass distribution is independent of the number of particles in the simulation. Nevertheless, a sufficiently high number of particles is required to correctly capture the dust dynamics near the planet and reduce the torque oscillations. We checked that increasing $N$ to 2 million particles per size bin does not significantly affect torque values (see Appendix~\ref{app:res_conv}).
Then, the mass distribution is converted to the dust surface density on the grid, $\Sigma _d (r, \varphi)$, using the cloud-in-cell method. From the surface density, we can evaluate the torques acting on the planet:
\begin{equation}
    \Gamma _{d,g} = \int _{\rm{disk}} \Sigma _{d,g} \frac{\partial \Phi (r, \varphi)}{\partial \varphi}\rm{d}S,
\end{equation}
where the subscript $d$ ($g$) indicates the dust (gas) torques, respectively. The (smoothed) gravitational potential for the planet is given by
\begin{equation}
    \Phi (r, \varphi) = - \frac{GM_{\rm{pl}}}{\sqrt{d^2 + \epsilon _{d,g}^2}},
\end{equation}
where $d$ is the distance of the point $(r, \varphi)$ from the planet and $\epsilon $ is the smoothing length. The latter is used to prevent singularities at the location of the planet and to account for the fact that we are approximating a three-dimensional distribution in 2D. For the gas, we set $\epsilon _g = 0.6H$, with $H = hr$ the pressure scale height, because this value reproduces vertical forces very well \citep{muller2012}, and the resulting gas torque is in agreement with 3D simulations \citep{dangelo2010, 2012ARA&A..50..211K}.  {In previous studies of the dust torque, \cite{regaly2025} used the same smoothing length, while \cite{benitez2018} and \cite{chrenko24} adopted a mass-dependent value ($\epsilon _g = 1 $ $R_{\rm{H}}$).} For dust particles, the choice of the smoothing length is less clear. Previous works have established that the main contribution to the dust torque comes from asymmetries in the dust distribution that originate very close to the planet \citep{benitez2018, regaly2020, chrenko24}: therefore, a small value of $\epsilon _d$ is required to have a significant number of dust particles approaching the planet. We also expect that dust particles will eventually accrete to the planet when they enter the Bondi sphere, $r_{\rm{B}} = GM_{\rm{pl}}/c _s^2$ \citep{2018MNRAS.479.5136P}. Therefore, we choose $\epsilon _d = 0$  {\citep{chrenko24}} and remove dust particles from the simulation when they cross $r_{\rm{B}}$, which avoids numerical divergences at the location of the planet.  {We note that the particles removed due to accretion are not re-initialized at the outer edge of the domain.} Setting the smoothing length to 0 is equivalent to assuming that all dust particles considered have settled on the midplane, which is reasonable considering the low level of turbulence and the evolved state of the disk, but it may not be correct for micrometer-sized grains. However, such particles are less massive and always tightly coupled to the gas; thus, we do not expect them to contribute significantly to the dust torque. 
 {It must be noted that the process of dust accretion cannot be resolved accurately with our code. However, we want to provide a simple prescription to identify particles that have likely become part of the circumplanetary environment and therefore should no longer contribute to the gravitational torque exerted on the planet.} Therefore, we also exclude dust grains inside the Hill sphere of the planet without sufficient kinetic energy to escape from it: $e_{\rm{kin}} + e_{\rm{grav}} < e_{\rm{grav}}(r_{\rm{H}})$ \citep{picogna2018}.  {On the one hand, if a dust particle is inside the Hill sphere and has a negative orbital energy with respect to the planet, it should not contribute to the torque exerted by the circumstellar disk; on the other hand, if a dust particle is inside the Hill sphere with a positive orbital energy, the gas flow will sweep it back to the surrounding disk.}  {As we discuss in Appendix~\ref{bound}, this energy-based criterion is only an approximation. Because gas drag is dissipative, particles may become temporarily bound before subsequently escaping the Hill sphere. In such cases, our algorithm may remove particles that would eventually return to the disk, leading to undesirable artificial perturbations in the dust density outside the Hill sphere. Nevertheless, comparison with alternative exclusion methods shows that this approximation has only a minor impact on the inferred dust torque for realistic maximum grain sizes (see \cifig{fig:bound}), while providing a physically motivated way to exclude dust that is likely to have been accreted onto the circumplanetary environment.}

Finally, we included turbulent diffusion of dust particles following \cite{charnoz2011}, incorporating periodic position jumps to simulate turbulent mixing. We modeled the kick on the particle position as a random Gaussian variable \citep[see][]{picogna2018}, with mean and variance depending on the dust diffusion coefficient, $D_d$, which is related to the gas diffusion
\begin{equation}
    D_d = \frac{D_g}{\rm{Sc}} \simeq \frac{\alpha c_s H}{\rm{Sc}},
\end{equation}
where to a first approximation $D_g \sim \alpha c_s H$ and Sc is the Schmidt number
\begin{equation}
\rm{Sc} = \frac{\left(1 + \Omega _{\rm{K}}^2 \tau _s ^2\right)^2}{1 + 4\Omega _{\rm{K}}\tau_s} = \frac{\left(1 + \rm{St} ^2\right)^2}{1 + 4\rm{St}}.
\label{diff}
\end{equation}
The last term shows that turbulent diffusion is more effective when the dust is tightly coupled to the gas.
At each time step and for every particle, a random number is generated and the probability of applying a position kick is the ratio $dt/\tau_c$ \citep[as in][]{YOUDIN2007588}, where $\tau_c$ is the correlation time, which is assumed to be equal to the eddy time $\tau_c \approx \tau_\mathrm{eddy} \approx 1/\Omega_K$ \citep{2006A&A...452..751F}.

\section{Results} \label{sec:results}

\begin{figure}
    \includegraphics[width=\linewidth]{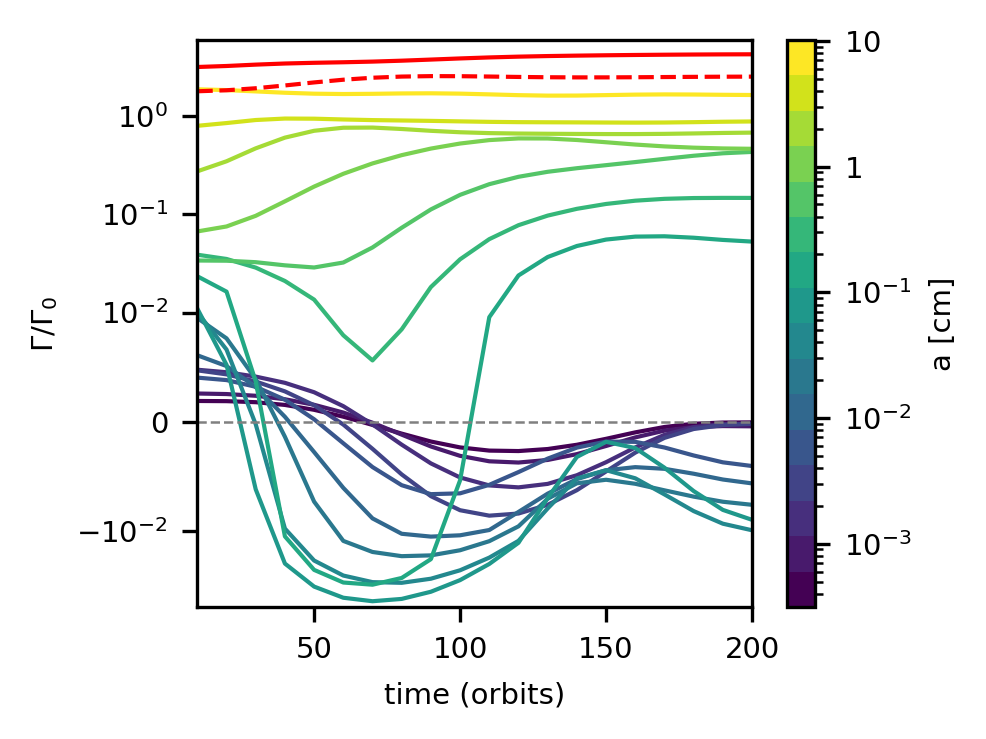}
    \caption{Torques acting on a $5\,M_{\oplus}$ planet as a function of time, normalized by $\Gamma_0$. The curves are smoothed over 10 orbits using a Gaussian window. Solid (dashed) red lines show the absolute value of the total dust (gas) torque, while the color scale shows the contribution from individual dust size bins.}
    \label{fig:torque_evo}
\end{figure}

\cifig{fig:torque_evo} shows the time evolution of the gas and dust torques for a model with a $M_{\rm pl} = 5\,M_{\oplus}$ planet embedded in a disk with $\Sigma_0 = 100$ g cm$^{-2}$. All torque values are normalized by

\begin{equation}
\Gamma_0 =
q^2
r_{\rm pl}^4
\left[
\Sigma_g \Omega_{\rm K}^2 h^{-2}
\right]_{r_{\rm pl}},
\label{norm}
\end{equation}
where $q = M_{\rm pl}/{M_*}$ is the planet-to-star mass ratio, $\Omega_{\rm K}$ is the Keplerian frequency and all quantities are evaluated at $r=r_{\rm pl}$.  

The torque is computed by recording the spatial distribution of gas and dust particles after each planetary orbit. The particle masses are then deposited onto the grid to reconstruct the dust surface density $\Sigma_d$ for each size bin using Eq.~\ref{mass}. The resulting torque is subsequently smoothed over  {10} planetary orbits using a Gaussian window.

As shown in \cifig{fig:torque_evo}, both gas and dust torques require approximately $150$ planetary orbits to reach a steady state. The resulting total torque is small but positive, implying slow outward migration. This behavior is consistent with previous studies \citep{benitez2018, regaly2020, chrenko24}.

\cifig{fig:torque_evo} also shows the contribution to the dust torque of each particle size bin. A key result is that only the five largest grain sizes contribute significantly to the total dust torque. Consequently, both the magnitude and the sign of the total torque—which determine the direction and rate of planetary migration—depend sensitively on the maximum grain size present in the disk.

\begin{figure*}[h!]
    \centering
    \includegraphics[width = \linewidth, keepaspectratio]{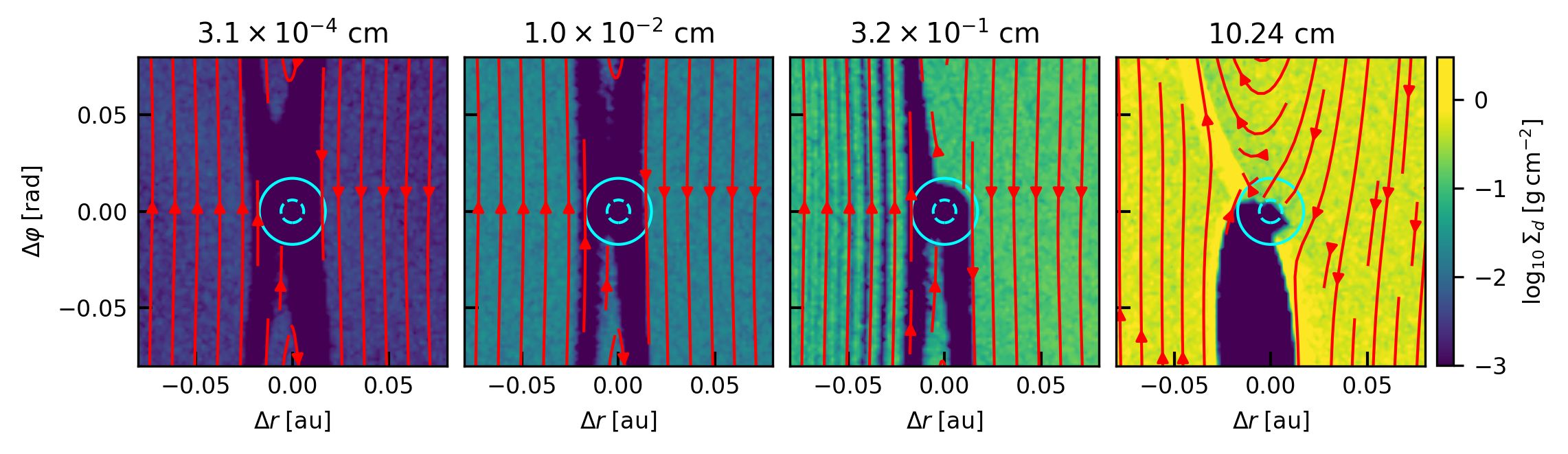}
    \caption{Dust densities for four representative dust sizes, overlaid with the particles' trajectories. The dust distribution is averaged over the last 50 orbits. Solid (dashed) circles indicate the Hill (Bondi) sphere of the planet, respectively.  {From left to right, the panels correspond approximately to St = $10^{-5}, 3 \times 10^{-4}, 10^{-2}$ and $ 5 \times 10^{-1}$}.}
    \label{fig:dust_dens}
\end{figure*}

Two main physical reasons explain this behavior. First, most of the dust mass is concentrated in the largest grains (see Eq.~\ref{mass}). Second, large particles develop a pronounced front–rear asymmetry in their surface density distribution relative to the planet. This effect is illustrated in \cifig{fig:dust_dens}, which shows the dust surface density $\Sigma_d$ for each size bin together with the corresponding particle trajectories.

In particular, a large dust-depleted region forms downstream of the planet for the $10.24$ cm grains. Similar structures have been reported in previous studies \citep{benitez2018, chrenko24} and arise from the scattering of weakly coupled particles by the planet.  In contrast, smaller grains remain strongly coupled to the gas and therefore closely follow the gas flow. Their gravitational interaction with the planet is correspondingly weaker, resulting in a much smaller front–rear asymmetry and a negligible contribution to the total torque (see \cifig{fig:torque_evo}).

In some cases the asymmetry is even reversed, meaning that the dust surface density is slightly lower ahead of the planet than behind it. This occurs for micrometer-sized grains that are tightly coupled to the gas, for which the drag force dominates the gravitational perturbation from the planet. As a result, these grains closely follow the gas spiral pattern and are only weakly scattered by the planet.  {The underdense patches observed in the distribution of small grains in \cifig{fig:dust_dens} may be related to perturbations of the gas flow produced by the planet. However, these patches are also partly affected by our exclusion criterion, which could remove dust particles that should return to the disk instead of being captured by the planet (see \cifig{fig:bound_density}).} 

 {As grain size increases, the particles are less coupled to the gas and are more prone to gravitational scattering by the planet, which} is particularly important for the torque calculation because particles passing close to the planet provide the dominant contribution to the dust torque.

However, this contribution can fluctuate rapidly as the particles cross the Hill sphere at high speed. For this reason, previous works introduced a radial cutoff when evaluating the torque \citep{benitez2018, regaly2020, chrenko24}. Instead, we provide a more physically motivated constraint by removing gravitationally bound particles, although this introduces slightly more oscillations to the instantaneous torque.

\begin{figure*}[h]
    \centering
    \includegraphics[width = \linewidth, keepaspectratio]{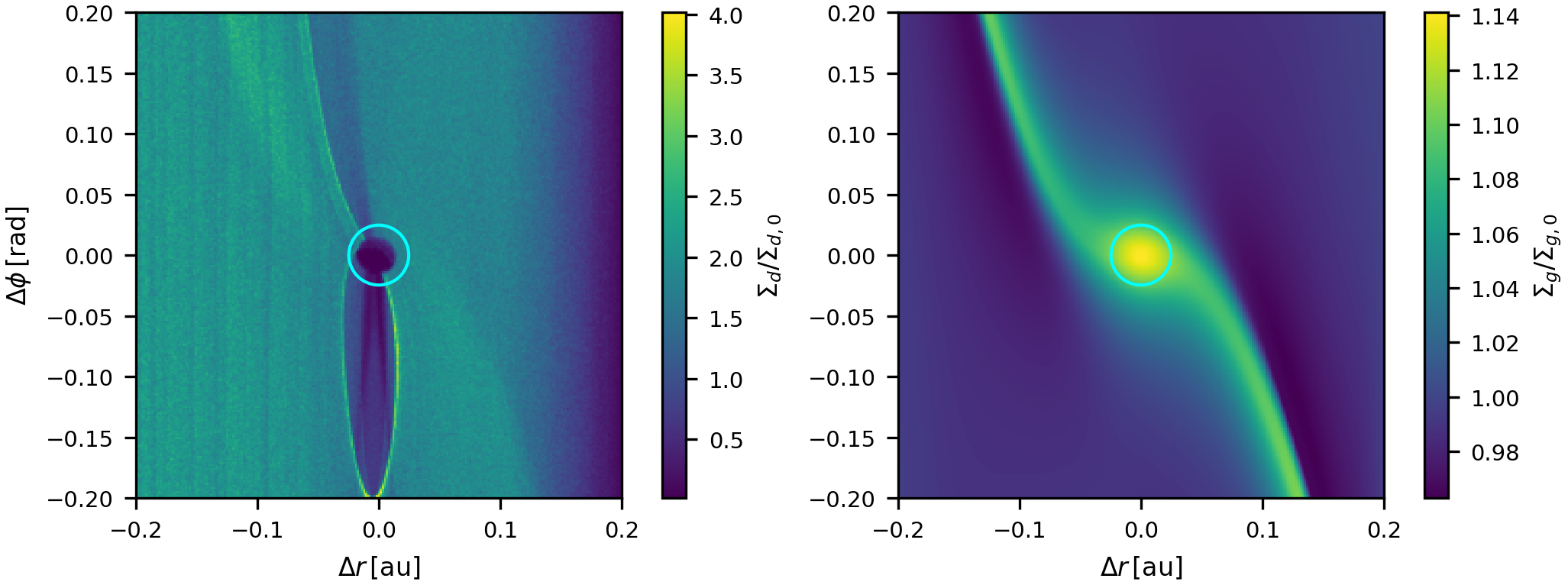}
    \caption{Dust and gas surface densities for the simulation with $M_{\rm{pl}} = 5 \, M_{\oplus}$ and $\Sigma _0 = 100$ g cm$^{-2}$. Circles indicate the Hill sphere of the planet.}
    \label{fig:total_dens}
\end{figure*}

Ultimately, the total dust torque on the planet is given by the total dust density, which represents the sum of the contribution from each dust size, weighted by their mass fraction (see Eq. \ref{mass}). We illustrate the (total) dust and gas surface densities in \cifig{fig:total_dens}. We note the spiral pattern followed by the strongly coupled grains, although it is less visible because of their lower mass fraction. In contrast, the weakly coupled pebbles deviate from the gas spiral because of the planet-induced scattering, which generates the dust hole behind the planet that is responsible for the positive torque, which is the dominant contribution because most of the dust mass is contained in the pebbles.  

\subsection{Dust torque as a function of $a_{\rm max}$}
Obtaining a realistic estimate of the dust torque requires knowledge of the maximum grain size present in the disk. Observationally, grain sizes are commonly inferred from spectral indices measured at millimeter wavelengths. However, converting spectral indices into grain sizes requires detailed radiative transfer modeling that accounts for the disk structure and the optical properties of the dust \citep{2024A&A...688A.204L}.

In our fiducial simulations we adopt $a_{\rm max} = 10.24$ cm, which is broadly consistent with the limited observational constraints available for the inner regions of protoplanetary disks. However, theoretical models predict that dust growth is limited by fragmentation, bouncing barriers, and radial drift \citep[see the review by][]{2024ARA&A..62..157B}. Moreover, \cite{2025ApJ...995L..19H} have shown that dust growth can be suppressed by planet-induced coagulation modes, thereby reducing the dust-driven torque.

\begin{figure}
\centering\resizebox{\linewidth}{!}{\includegraphics[clip]{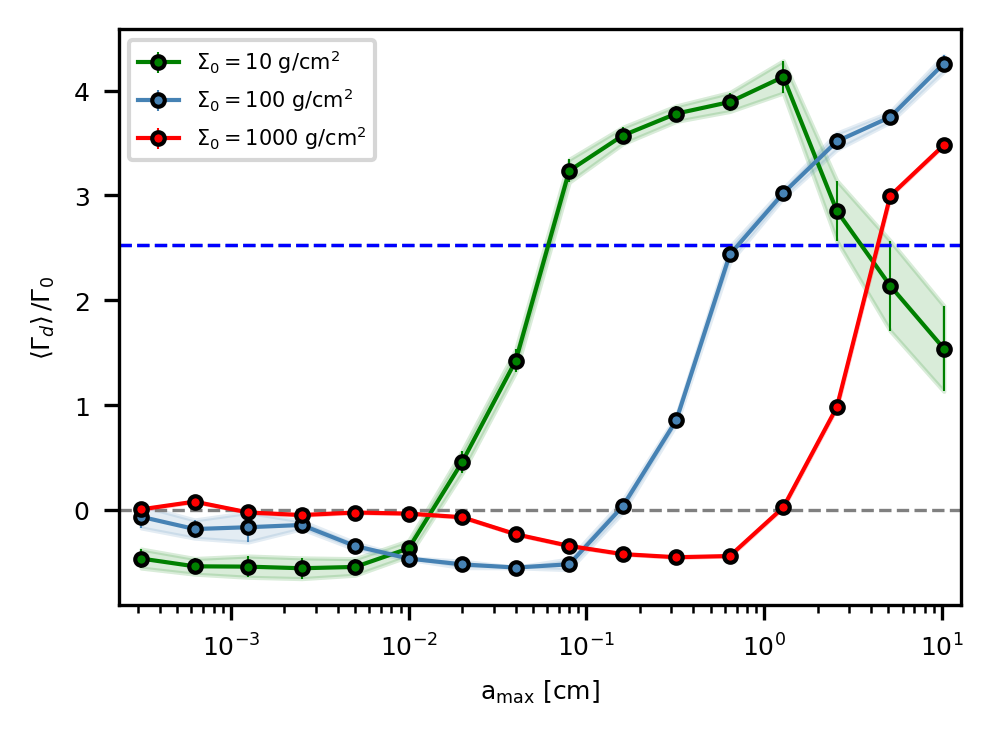}}
\centering\resizebox{\linewidth}{!}{\includegraphics[clip]{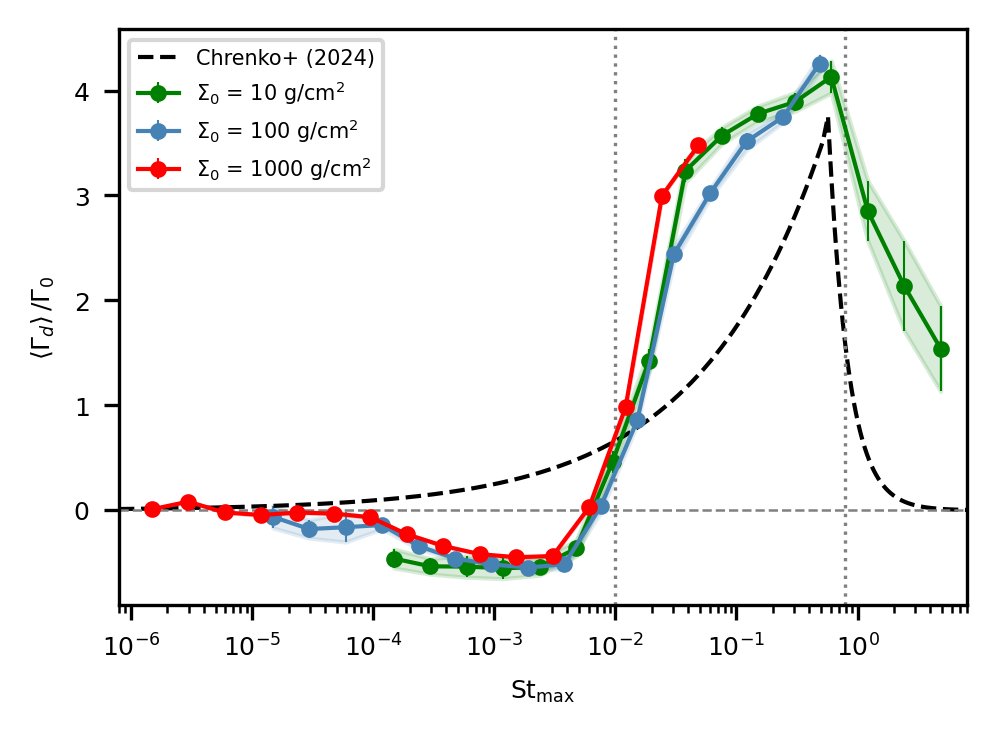}}
    \caption{Top: total dust torque averaged over the last 10 orbits as a function of the maximum grain size,  for three different simulations with varying gas surface density ($\Sigma _0$). Blue dashed line shows the absolute value of the gas torque, that is the same in all simulations. Bottom: same result as a function of the maximum Stokes number. Black dashed line shows the scaling law proposed by \cite{chrenko24, 2026A&A...706C...1C}, with dotted lines corresponding to the minimum and maximum St used in their study.}
    \label{fig:torque_amax}
\end{figure}

To investigate how the dust torque varies with the maximum grain size, we retain the same mass distribution given in Eq.~\ref{mass} but truncate it at different values of $a_{\rm max}$. This approach allows us to remain agnostic about the exact value of $a_{\rm{max}}$, given the contrast between theoretical predictions and observations. The total resulting dust torque as a function of $a_{\rm max}$ is shown in \cifig{fig:torque_amax}.

In the simulation with $\Sigma _0 = 100$ g cm $^{-2}$, the dust torque becomes positive for $a_{\rm max} \gtrsim 0.1$ cm and exceeds the gas torque for $a_{\rm max} \gtrsim 5$ cm, implying that the outward migration of a $M_{\rm pl} = 5\,M_{\oplus}$ planet requires the presence of relatively large pebbles. Since such large grains are expected to be limited by fragmentation and drift, sustained outward migration driven by dust alone may be difficult to achieve under realistic conditions. We show in the bottom panel of \cifig{fig:torque_amax} our results for the total dust torque as a function of the Stokes number of the particles. The results are consistent with the torque scaling law with fixed Stokes number proposed by \cite{chrenko24} (later corrected in \cite{2026A&A...706C...1C}), at least for the values of St explored in their work (St $\in [10^{-2}, 0.785]$).
Deviations from this scaling are expected due to multiple effects:

\begin{itemize}
    \item Small dust grains that are strongly coupled to the gas dynamics ($\rm{St} \lesssim 10^{-3}$) tend to follow the gas spiral rather than being scattered by the planet. As a result, the front-rear asymmetry near the planet is significantly modified (see \cifig{fig:dust_dens}), reducing the torque amplitude and, in some cases, reversing its sign. A similar trend is reported by \cite{2024ApJ...972..152H} with a linear analysis of the dust torque. In this regime, the planet would migrate inward more rapidly as dust adds a negative contribution on top of the already negative gas torque. 

    \item Because we consider a dust size distribution rather than a single and constant Stokes number, each grain size contributes with a different weight to the total torque. Nevertheless, the largest grains present in the disk ($a_{\rm max}$) dominate the torque because they contain most of the dust mass and are weakly coupled to the gas.

    \item Instead of applying a radial cutoff to exclude the Hill sphere from the torque calculation, we remove from the simulation the dust particles that become gravitationally bound to the planet. Roughly speaking, these particles should be considered as part of the planet and should therefore not contribute to the torque. However, large grains can enter the Hill sphere at high velocity and exert a significant torque on the planet before escaping from its gravitational influence.
\end{itemize}

The torque dependence on $a_{\rm max}$ is highly sensitive to the disk properties, particularly the gas surface density. The case with $\Sigma_0 = 100$ g cm$^{-2}$ represents the most significant scenario for dust-driven torque, since the largest grains contain most of the dust mass and develop the strongest asymmetry near the planet. In \cifig{fig:torque_amax}, we also show results for a lower-density disk ($\Sigma_0 = 10$ g cm$^{-2}$) and a higher-density disk ($\Sigma_0 = 1000$ g cm$^{-2}$), which effectively shifts the grain population towards higher or lower Stokes numbers, respectively.  

In the low-density disk, the dust torque becomes positive for $a_{\rm max} \gtrsim 10^{-2}$ cm, reaches a maximum near $a_{\rm max} \sim 1$ cm, and subsequently decreases for larger grain sizes. This decrease is caused by the increase in the radial drift velocity of weakly coupled particles. When $\mathrm{St} \gtrsim 1$, grains traverse the planetary region on a timescale shorter than the characteristic gravitational interaction time, entering a quasi-ballistic regime. In this limit, the planet is unable to efficiently deflect pebble trajectories, and the dust distribution around the planet becomes less coherent. Therefore, the front-rear asymmetries responsible for the positive torque are strongly reduced.

As a consequence, the contribution of the largest pebbles to the total torque becomes negligible, despite their dominant share of the dust mass. This behavior differs from the trends reported in non-accreting simulations \citep{benitez2018, 2026A&A...706L...7G}, but is consistent with the scaling proposed by \cite{chrenko24}, in which the dust torque rapidly decreases for $\mathrm{St} \gtrsim 1$ (see \cifig{fig:torque_amax}). Nevertheless, we still measure a substantial positive dust torque owing to the contribution from intermediate-sized grains, which remain sufficiently massive while still interacting efficiently with the planet. This result highlights the importance of adopting a realistic dust size distribution rather than assuming a single, fixed Stokes number.

In the high-density case, the dust torque is positive only for $a \gtrsim 1$ cm, becomes negative for millimeter-sized grains, and tends to zero for micrometer-sized particles. This behavior can be attributed to the stronger coupling between dust and gas, together with the enhanced effect of turbulent diffusion, which smooths the dust distribution and suppresses the front–rear asymmetries responsible for the torque. Such conditions are representative of younger, more massive disks, where large grains are more likely to be present. However, even in this case, the dust torque remains below the gas torque, so the total torque is negative and migration is directed inward. Nevertheless, the contribution of pebbles can significantly reduce the net torque, potentially slowing down or even temporarily halting the inward migration. 

\begin{figure}
    \centering
    \includegraphics[width=\linewidth]{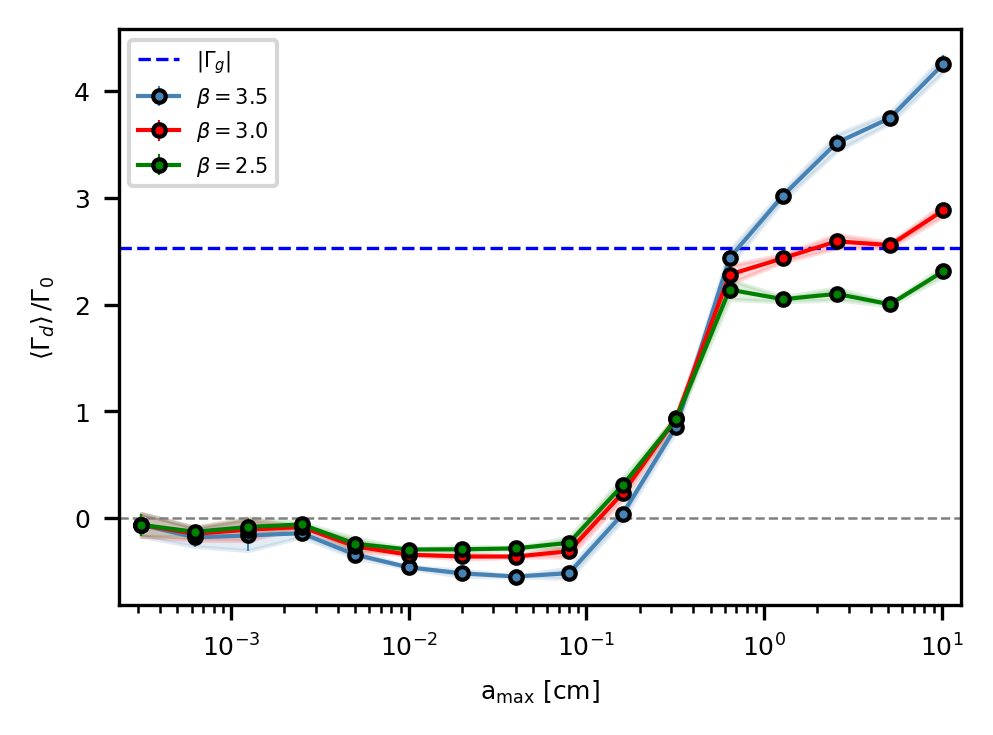}
    \caption{Total dust torque for the simulation with $M_{\rm{pl}} = 5 \, M_{\oplus}$, $\Sigma _0 = 100 $ g cm $^{-2}$, and three different values of the slope of the dust size distribution.}
    \label{fig:torque_n}
\end{figure}

Another factor that can affect the dust torque is the slope of the grain size distribution $\beta$. We adopt $\beta = 3.5$ as our standard, which corresponds to the canonical distribution for dust particles in the interstellar medium and in collisional cascades, such as debris disks \citep{MRN}. However, in protoplanetary disks, dust coagulation plays a fundamental role in shaping the size distribution \citep{2012A&A...539A.148B}. The system reaches a fragmentation-coagulation equilibrium, where the particles grow until they reach a size at which they break up. In this regime, the slope of the size distribution depends on the dominant source of relative velocities—whether fragmentation or radial drift \citep{2024ARA&A..62..157B}.  

Models of dust evolution indicate that drift-limited distributions are often “top-heavy,” with a flatter slope ($\beta \sim 2.5$) in the upper part of the distribution \citep{2024ARA&A..62..157B}. In \cifig{fig:torque_n}, we show the total dust torque for three different values of $\beta$, for a planet of $M_{\rm pl} = 5 \, M_{\oplus}$ embedded in a disk with $\Sigma_0 = 100$ g cm$^{-2}$. The results indicate that the torque is sensitive to the slope mainly for the largest grains. For flatter distributions ($\beta \lesssim 3$), the number of pebbles in the disk is reduced, and thus the total dust torque remains below the gas torque for all grain sizes, implying that planetary migration would always be directed inward.

It is important to note that the spatial distribution of small grains evolves on longer timescales than that of larger particles. As discussed by \cite{chrenko24}, the torque can vary significantly on a timescale comparable to the horseshoe crossing time due to radial drift. This behavior is visible in \cifig{fig:torque_evo}, where the torque peak is shifted to later times as the grain size decreases. Our simulations are limited to $\sim 200$ orbits, which is similar to the crossing time for intermediate Stokes numbers ($\mathrm{St} \sim 3 \times 10^{-2}$); this explains the sudden increase in the dust torque seen in \cifig{fig:torque_amax} at $a_{\rm{max}} = [0.08, 0.64, 5.12]$ cm for $\Sigma _0 = [10, 100, 1000]$ g cm$^{-2}$, respectively. 

However, for smaller grains, the drift velocity is much lower and the corresponding crossing time can reach $t_{\rm cross} = 2x_s /|u_r| \sim 5 \times 10^{5}\,P_{\rm orb}$, where $x_s$ is the half-width of the horseshoe region and $u_r$ the drift velocity \citep{chrenko24}. Extending the simulations over such long timescales would introduce additional complications. First, the planet would undergo significant migration on a timescale $\tau _{\rm{mig}} = L_{\rm{pl}}/2\Gamma \sim t_{\rm{cross}}$, with $L_{\rm{pl}} = M_{\rm{pl}}\sqrt{GM_{*}r_{\rm{pl}}}$ the angular momentum of the planet (which is kept fixed in our setup). Second, the disk itself would evolve on a viscous timescale $\tau_\nu = r^2/\nu \sim t_{\rm{cross}}$ for our disk parameters, where $\nu = \alpha c_s H$ is the kinematic viscosity \citep{2020apfs.book.....A}. Nevertheless, in \cifig{fig:torque_evo} we observe that the residual torque variations remain within a factor of $\lesssim 2$.

One caveat of our model is that we neglect the dust backreaction on the gas. This has been shown to enhance positive torques and reduce the magnitude of negative torques, especially when pebble accretion is included \citep{regaly2025}. Since dust backreaction is expected to be strongest for weakly coupled particles, its inclusion would likely amplify the contribution from the largest grains and modify the gas response. As a result, both the magnitude of the dust torque and its dependence on $a_{\rm max}$ may be steeper than reported here. However, the situation may become more complex once additional physical effects are included. In particular, \cite{2025A&A...704A.207C} found that in non-isothermal 3D disks, dust backreaction can lead to the formation of asymmetric dust overdensities analogous to the cold thermal lobes that develop in the gas component. In their non-accreting simulations, the dust torque becomes increasingly negative with time. Future work should address this complex interplay between dust backreaction, accretion, and heating or cooling processes, which could change considerably the dust torque.

\subsubsection{Dependence on planetary mass}

\begin{figure}
    \centering
    \includegraphics[width=\linewidth]{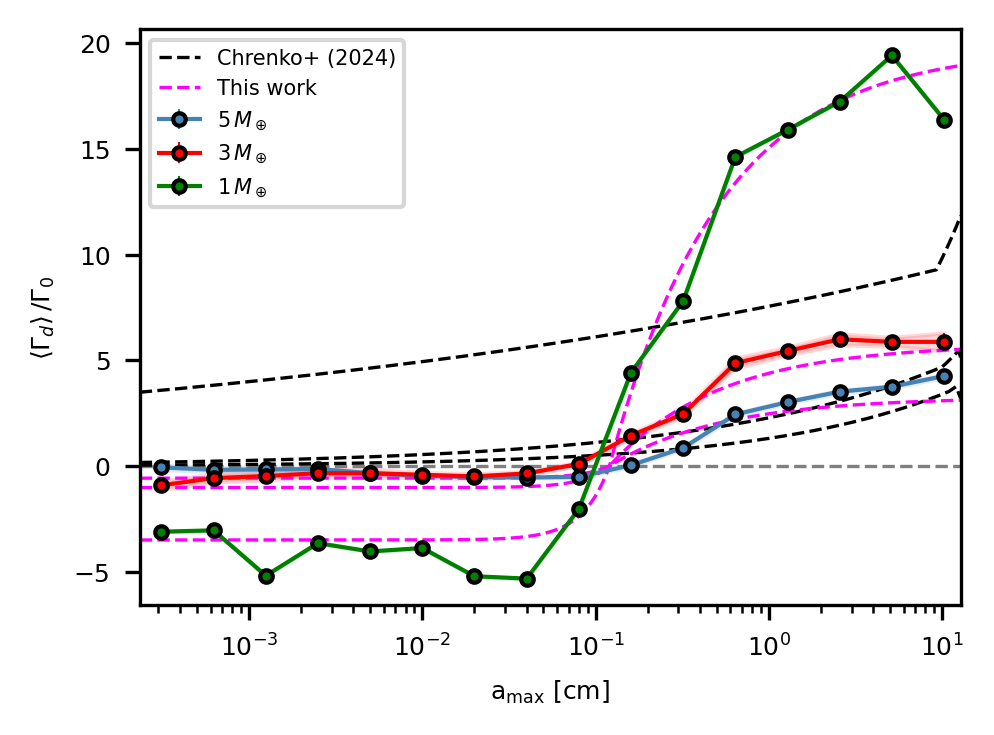}
    \caption{Same as \cifig{fig:torque_amax} but for three different planetary masses and $\Sigma _0 = 100$ g cm$^{-2}$.}
    \label{fig:torque_mass}
\end{figure}

In Eq.~\ref{norm}, the torque is normalized by the square of the planet-to-star mass ratio, as commonly done for gas torques \citep{dangelo2010}. However, scaling of the dust torque with planetary mass is not expected to follow the same behavior. In particular, previous studies have shown that the relative importance of dust torque increases for Earth-mass planets \citep{benitez2018, chrenko24}.

Exploring the dependence of the dust torque on the planetary mass is numerically challenging. As the planetary mass decreases, the Hill sphere shrinks, requiring higher spatial resolution to accurately resolve particle trajectories in the vicinity of the planet. At the same time, close encounters between dust particles and the planet become increasingly important, introducing strong nonlinearities in the evaluation of the torque. As mentioned above, we chose the resolution to have approximately 10 squared cells per Hill radius (702x7350 for $M_{\rm{pl}} = 1 \, M_{\oplus}$ and 417x4365 for $M_{\rm{pl}} = 3 \, M_{\oplus}$).

The resulting dust torque, averaged over the last 10 orbits, is shown in \cifig{fig:torque_mass}. Compared to the $5 \, M_{\oplus}$ case, we find substantial differences. In the presence of pebbles ($a_{\rm max} \gtrsim 1$ cm), the dust torque significantly exceeds the gas torque, which is generally consistent with the scaling proposed by \cite{chrenko24}. In contrast, when the maximum grain size is small, the dust torque becomes more negative, especially for an Earth-mass planet. Interestingly, the maximum grain size at which the dust torque shifts from negative to positive values, $a_0$, is roughly the same for each planetary mass. We can identify two different regimes: when $a_{\rm{max}} \gtrsim a_{0}$, the gravitational interaction with the planet dominates the gas drag force: as a result, the trajectories of dust grains deviate significantly from the gas spiral, generating an asymmetry around the planet that exerts a positive torque. In contrast, for $a_{\rm{max}} \lesssim a_{0}$, the grains are tightly bound to the gas dynamics, resulting in a negative dust torque on the planet. Given these results, we propose a new empirical scaling law for the dust torque with the maximum dust size in the disk, which could be implemented in population synthesis studies:

\begin{equation}
\frac{\Gamma_d}{\Gamma_0}
= \frac{Z}{0.01}\left(\frac{q}{1 \,M_{\oplus}}\right)^{-1.12}e^{0.2+0.5(\beta-3.5)} f(a_{\rm max}),
\end{equation} \label{fit}
where  {$Z$ is the dust to gas ratio, $\beta$ is the slope of the dust size distribution and }the size-dependent function $f(a_{\rm max})$ is given by
\begin{equation}
f(a_{\rm max}) =
\begin{cases}
-3.49 \left[1 - \exp\left(-6.43\log _{10}{\frac{a_0}{a_{\rm max}}}\right)\right],
& a_{\rm max} \le a_0, \\[6pt]
\;\;19.77 \left[1 - \exp\left(-1.55\log _{10}{\frac{a_{\rm max}}{a_0}}\right)\right],
& a_{\rm max} > a_0,
\end{cases}
\end{equation}
with $a_0 \simeq 0.12\,\mathrm{cm}$. The function $f(a_{\rm max})$ also depends on the local gas surface density ($\Sigma_0 = 100$ g cm$^{-2}$ in our fiducial model), although a similar scaling is recovered when expressed in terms of the maximum Stokes number, $\mathrm{St}_{\rm max}$ (see \cifig{fig:torque_amax}).  {In Eq. \ref{fit}, the planet-to-star mass ratio $q$ is expressed in units of Earth masses to emphasize that we only explored values of $M_{\rm{pl}} = 1 - 10$ $M_{\oplus}$, so that the fit may not be appropriate for planetary masses outside of this range. In particular, our} scaling law is valid for planetary masses below the pebble isolation mass, $M_{\rm iso} \sim 20 \, M_{\oplus}$ for our adopted disk parameters \citep{2018A&A...612A..30B}. For $M_{\rm pl} \gtrsim M_{\rm iso}$, the planet generates a pressure bump that traps drifting pebbles outside of its orbit, preventing close encounters with the planet and effectively suppressing the dust torque. \cite{2018A&A...612A..30B} show how $M_{\rm{iso}}$ scales with the viscosity of the disk, the Stokes number of the dust particles, and the aspect ratio of the disk. In a cold disk, the pebble isolation mass is significantly reduced ($M_{\rm iso} \sim 4 \, M_{\oplus}$ for $h = 0.03$), limiting the regime in which dust torques can be effective. However, if $h \gtrsim 0.04$, $M_{\rm{iso}}$ is always above $5 \, M_{\oplus}$, for any value of viscosity and Stokes number. Therefore, if the disk cools in the midplane or the planet accretes enough mass to reach $M_{\rm{pl}} \sim M_{\rm{iso}}$, we expect the contribution of the dust torque to vanish and thus the planet will migrate inward, as predicted by Type I theory \citep{2002ApJ...565.1257T}.

\subsubsection{Dependence on viscosity}

\begin{figure}
    \centering
    \includegraphics[width=\linewidth]{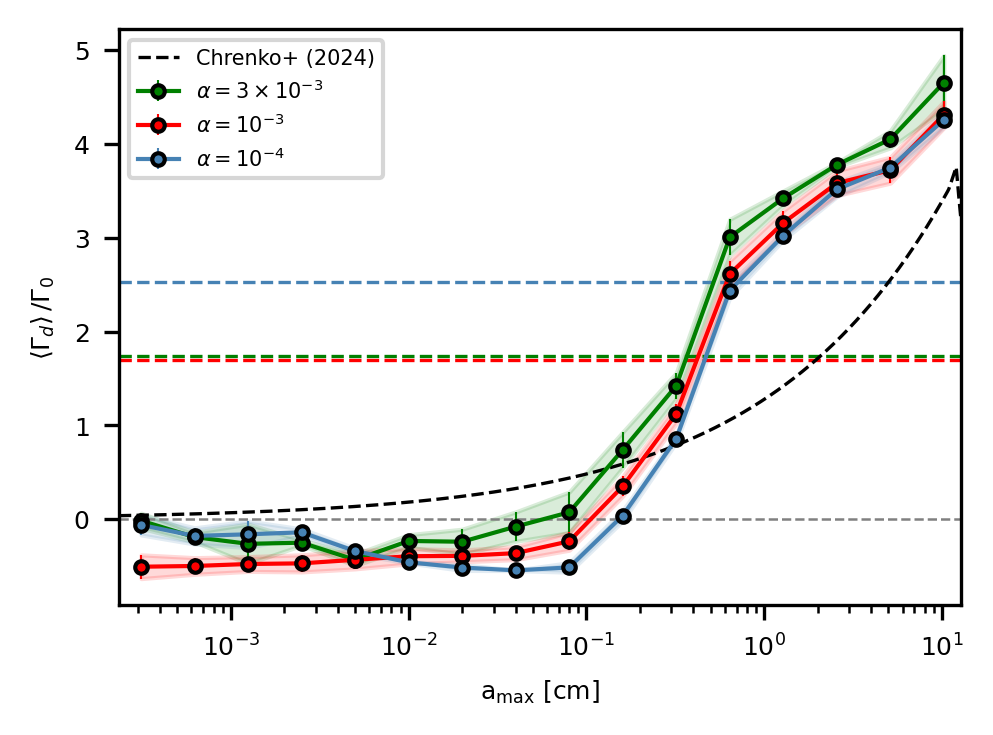}
    \caption{Same as \cifig{fig:torque_amax} but for three different values of the viscosity parameter $\alpha$. Colored dashed lines show the corresponding absolute value of the gas torque.}
    \label{fig:torque_visc}
\end{figure}

\begin{figure*}[h!]
    \centering
    \includegraphics[width = \linewidth, keepaspectratio]{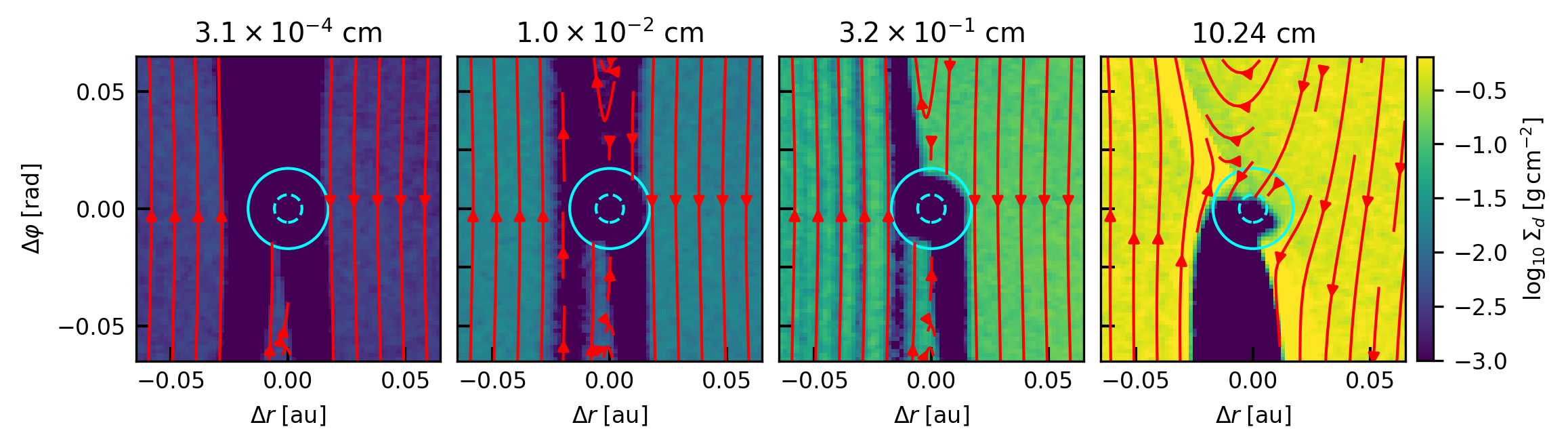}
    \caption{ {Same as \cifig{fig:dust_dens}}, but for $\alpha = 3 \times 10^{-3}.$}
    \label{fig:dust_dens_visc}
\end{figure*}

We performed additional simulations with different values of $\alpha$, to check the influence of the viscosity parameter on the dust torques. In our model, increasing $\alpha$ is equivalent to increasing the dust diffusion coefficient (see Eq. \ref{diff}). This has been reported to decrease the dust torques for a fixed Stokes number by \cite{2025ApJ...979..185H} and \cite{2025A&A...698A..21C}, because diffusion is expected to smooth the dust density gradients and reduce the asymmetries near the planet that are responsible for the torque. However, \cite{2025A&A...698A..21C} also find that the dust-driven migration of planets with mass $M_{\rm{pl}} \sim 3-12 \, M_{\oplus}$ is directed outward and is faster for increased $\alpha$. In \cifig{fig:torque_visc} we show the results for our fiducial simulation ($M_{\rm{pl}} = 5 \, M_{\oplus}$, $\Sigma _0 = 100$ g cm$^{-2}$) and three different values of $\alpha = [10^{-4}, 10^{-3}, 3 \times 10^{-3}]$. We find a slight increase of the dust torque with $\alpha$ for all dust sizes, while the gas torque decreases in absolute value (dashed lines in \cifig{fig:torque_visc}). This may be related to the positive contribution given by the corotation torque, which for large $\alpha$ behaves as expected from linear theory, while saturates at low $\alpha$, leaving only the linear Lindblad torques \citep{2009MNRAS.394.2283P}. In this context, the effect of viscosity is to restore the original gas density profile.  {However, dust density perturbations persist even for $\alpha = 3 \times 10 ^{-3}$, as shown in \cifig{fig:dust_dens_visc}. This differs from previous studies of turbulent dust diffusion \citep{2025A&A...698A..21C,2025A&A...704A.207C}, possibly because our simulations include an approximate treatment of pebble accretion, which was neglected in those works. Another possible explanation is that we neglect the dust back-reaction on the gas, although \cite{regaly2025} has shown that it can enhance the effect of dust torques when combined with pebble accretion. Finally,} it must be noted that with sufficiently low viscosity even a low-mass planet could open a dust gap \citep{2025A&A...703A.270R} and change the resulting dust torque. Therefore, the effect of viscosity is potentially degenerate with that of planetary mass and may require a larger exploration of the parameter space.

\subsection{Torque profiles}

\begin{figure}
    \centering
    \includegraphics[width = \linewidth]{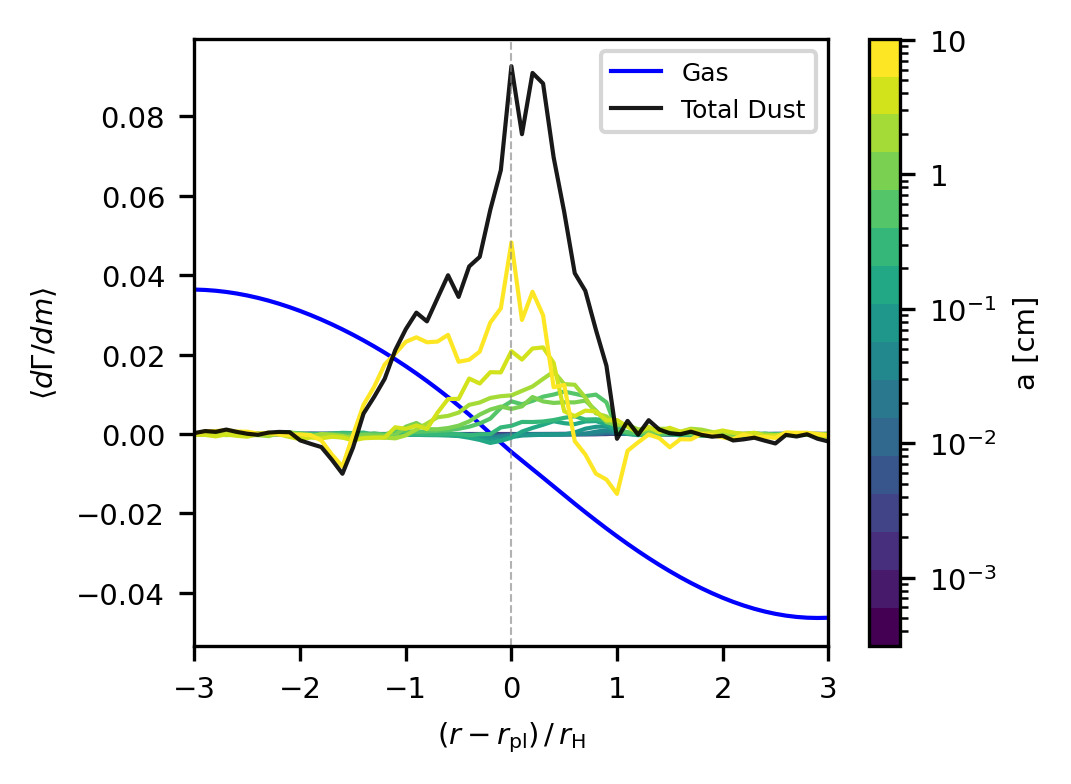}
    \caption{Azimuthally averaged torque density as a function of radial distance from the planet with mass $M _{\rm{pl}} = 5 \, M_{\oplus}$, expressed in units of Hill radii.}
    \label{fig:torque_profile}
\end{figure}

We investigate the radial distribution of the dust torque, considering the contribution of each dust size bin. \cifig{fig:torque_profile} shows the azimuthally averaged torque density as a function of the radial distance from the planet, expressed in units of the Hill radius.

Following \cite{dangelo2010} (their Eq.~14), we normalize the torque per unit mass as
\begin{equation}
    \frac{{\rm d}\Gamma_0}{{\rm d}M}(r) =
    \Omega_{\rm K}^2(r_{\rm pl})\, r_{\rm pl}^2
    \left(\frac{M_{\rm pl}}{M_*}\right)^2
    \left(\frac{r_{\rm pl}}{H}\right)^2.
\end{equation}

The gas torque is distributed over a relatively large region, extending approximately $\pm 10\,r_{\rm H}$ from the planet, in agreement with \cite{regaly2020}. As expected, the inner disk exerts a positive torque, while the outer disk contributes negatively, resulting in an asymmetric profile and a net negative torque that drives inward migration. Because dust backreaction is neglected, the gas torque remains unaffected by dust dynamics.

In contrast, the dust torque is highly localized. The dominant contribution arises from within the Hill sphere and becomes negligible beyond $\sim \pm 2\,r_{\rm H}$. The radial profile exhibits a sharp peak near the planet, followed by a rapid decline with distance. Unlike the gas, both the inner and outer disk contribute predominantly positive torques, reflecting the accumulation of dust ahead of the planet and the formation of a depleted region in its wake. This behavior emphasizes that an accurate treatment of dust dynamics in the immediate vicinity of the planet is essential to reliably capture the total dust torque, whereas the gas torque originates from a broader region and is therefore less sensitive to local resolution.

In our simulations, a large number of particles enter the Hill sphere because the planetary potential acting on the dust is not smoothed. As a consequence, the torque contribution is dominated by close encounters with the planet. In this regime, numerical accuracy becomes critical: higher-order integrators could in principle provide a more accurate description of particle trajectories within the Hill sphere, although at a significantly increased computational cost. However, this would likely reduce the oscillations in the instantaneous torque measurements without significantly changing the dominant trend.

The torque profile also shows a strong dependence on grain size. In particular, the five largest sizes ($a \gtrsim 6.4$ mm) account for approximately $97\%$ of the total dust torque, and the largest grains ($a \sim 10$ cm) alone contribute approximately $46\%$. This reflects both their larger mass fraction and their stronger dynamical decoupling from the gas, which enhances the asymmetry of their spatial distribution around the planet.

\subsection{Migrating planets}

\begin{figure}
\centering
\includegraphics[width=\linewidth]{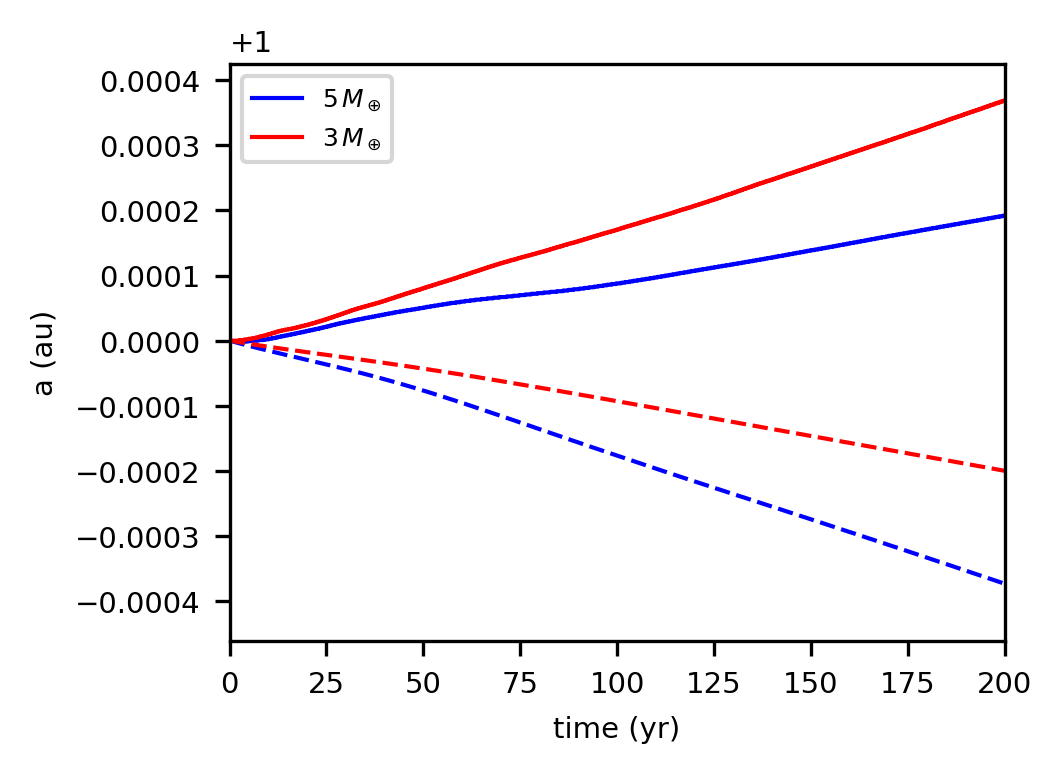}
\caption{Temporal evolution of the semimajor axis of planets with masses $M_{\rm pl} = 3$–$5\,M_{\oplus}$ embedded in a purely gaseous disk (dashed lines) and in a dusty disk (solid lines).}
\label{fig:migration}
\end{figure}

We now allow the planets to migrate freely instead of keeping them on fixed orbits. For the adopted parameters ($\Sigma_0 = 100$ g cm$^{-2}$, $\alpha = 10^{-4}$, $r_{\rm pl} = 1$ au, and $M_{\rm pl} = 3$–$5\,M_{\oplus}$), migration is expected to occur in the Type I regime. Due to the relatively low surface density and viscosity, the gas-driven migration rate is modest but increases approximately linearly with the mass of the planet \citep{2002ApJ...565.1257T, 2016JGRE..121.1962M}.

In contrast, the dust-driven contribution becomes increasingly important for lower-mass planets (see Eq.~\ref{fit}). When pebbles are present, the positive dust torque can exceed the negative gas torque, potentially reversing the direction of migration. In \cifig{fig:migration}, we compare the evolution of planets embedded in a purely gaseous disk (dashed lines) with that of a dusty disk that includes grains up to $a_{\rm max} = 10.24$ cm (solid lines).

For the planetary masses considered here, the semimajor axis evolves only slightly over the duration of the simulations, as the setup was designed to keep the planet within the computational domain while still capturing its dynamical feedback on the disk. In contrast to \cite{2020MNRAS.497.2425H}, we do not observe significant differences in the dust distribution between migrating and fixed-planet runs, most likely because the migration rate is too low to modify the local dust dynamics during the integration time. As a result, the torques measured in the migrating simulations remain essentially consistent with those obtained in the fixed-orbit case. We can assume, as a first approximation, that the migration rate remains constant over longer timescales, since both the dust and gas torques have reached a steady state (see \cifig{fig:torque_evo}). Under this assumption, a clear divergence emerges: in a purely gaseous disk, the planet would migrate inward and fall onto the central star on a timescale of the order $\sim 1$ Myr. In contrast, when dust—particularly in the form of pebbles—is included, the total torque becomes positive, leading to outward migration at a comparable rate. In this case, the planet would instead move outward to $\sim 2$ au over $\sim 1$ Myr, potentially surviving disk dispersal and growing towards a giant planet.

\section{Conclusions} \label{sec:conclusions}
We have investigated the torque exerted by a dusty protoplanetary disk on an embedded low-mass planet, using 2D hydrodynamical simulations with dust treated as Lagrangian superparticles. Our aim was to characterize the dust torque resulting from a realistic grain size distribution ranging from micrometer-sized particles to centimeter-sized pebbles. We find that both the magnitude and the sign of the dust torque are mainly determined by the largest grains present in the disk and their degree of coupling to the gas (see \cifig{fig:torque_amax}).

For a $5 \, M_{\oplus}$ planet, we find a positive dust torque when the maximum grain size corresponds to marginally coupled particles ($\mathrm{St}_{\max} \gtrsim 10^{-2}$). In this regime, particle trajectories are significantly perturbed when approaching the planet's Hill sphere, leading to a pronounced azimuthal asymmetry that drives a positive torque. When St $\gtrsim 1$, the drift speed of the pebbles is too high for the planet to significantly perturb their trajectories. As a consequence, their contribution to the torque is negligible, but the total dust torque remains positive due to the presence of intermediate-size particles. In the case of stronger coupling ($10^{-4} \lesssim \mathrm{St} \lesssim 10^{-2}$), the dust mimics the dynamics of the gas, resulting in a negative contribution to the torque. When St $\lesssim 10 ^{-4}$, the combination of strong dust-gas coupling and turbulent diffusion tends to cancel the asymmetries. Therefore, dust torques can drive outward migration of a super Earth only in the presence of sufficiently large pebbles ($a_{\rm{max}} \gtrsim 5$ cm  for $\Sigma _g \sim 100$ g cm$^{-2}$). Such conditions are likely difficult to maintain in real disks, where grain growth is limited by fragmentation and radial drift. Nevertheless, dust can contribute to slowing down the inward migration rate compared to the usual Type I prediction.

In contrast, for an Earth-mass planet, the dust torque is significantly enhanced, exceeding the gas torque by a factor of $\sim 3$–$4$, depending on the maximum grain size. This behavior is primarily related to the smaller Hill sphere, which reduces the fraction of particles that become gravitationally bound and instead allows a larger number of grains to undergo close encounters with the planet. In particular, pebbles approach the planet mostly from the front, exerting a strong positive torque that can lead to fast outward migration. The implication is that Earth-mass embryos would migrate outward rapidly, potentially reaching $\sim$10 au on disk lifetimes — placing them in the giant planet formation zone. This can have important consequences on current models of Solar system formation \citep{2021NatAs...5..898B}.

At the same time, the negative torque exerted by small (micrometer-sized) grains becomes more pronounced. Such particles remain tightly coupled to the gas and therefore follow the gas spiral, rather than being scattered by the planet. However, their contribution to the total dust torque remains limited, as their mass fraction is small compared to that of larger grains, which generate the most significant asymmetries and dominate the overall torque budget when present in the disk.

We also show how the dust torque varies with the assumed slope of the size distribution (see \cifig{fig:torque_n}). Flatter distributions reduce the number of pebbles in the disk and thus the total torque. Moreover, the torque density radial profile shows how most of the contribution comes from within the Hill sphere of the planet, emphasizing the need for high spatial resolution and precise numerical methods to accurately determine the effect of dust on the migration history of planets.

Finally, we provide a new scaling law for the dust torque, which incorporates the dependence on the planetary mass and the maximum dust size in the disk. This empirical formula can be implemented in a global planet formation model coupled with a dust evolution prescription \cite{2023ApJ...953...97G, 2025ApJ...986..199G}, which would allow the first self-consistent population synthesis including grain-size-dependent migration.

\begin{acknowledgements}
 {We thank the anonymous referee for their helpful and constructive comments, which greatly improved the clarity and robustness of our work.} This publication was produced while attending the PhD program in Astronomy at the University of Padova, Cycle XXXIX, with the support of a scholarship co-financed by the Ministerial Decree no. 118 of 2nd March 2023,760 based on the NRRP funded by the European Union - NextGenerationEU - Mission 4 Component 1 -- CUP C96E23000340001. 
\end{acknowledgements}

\bibliographystyle{bibtex/aa.bst}
\bibliography{local}

\begin{appendix}
\section{Grid resolution convergence}\label{app:res_conv}
\begin{figure}[h]
    \centering
    \includegraphics[width=\linewidth]{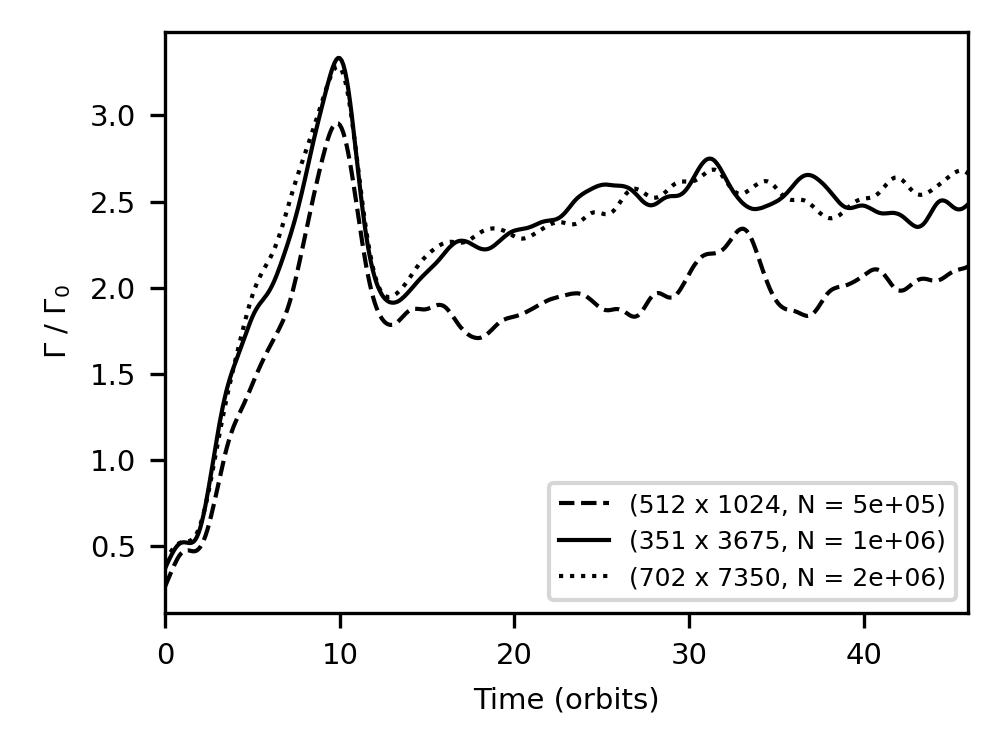}
    \caption{Numerical convergence test using three different resolutions, indicated in the legend. The figure shows the time evolution of dust torque felt by a 5 M$_{\oplus}$ planet due to a single dust size $a = 10.24$ cm. In this test we have no energy constraint on dust particles.}
    \label{fig:resolution}
\end{figure}

We test that the hydrodynamical solutions are in the numerically convergent regime. We perform simulations with $\rm{M}_{\rm{pl}} = 5  $ M$_{\oplus}$ and $\Sigma _0 = 100 $ g/cm$^2$, using three different resolutions ($N_{\rm{R}} \times N_{\varphi} = [512 \times 1024, 351 \times 3675, 702\times 7350]$) with increasing total number of dust particles $N = [5 \times 10^5, 1 \times 10^6, 2 \times 10^6]$ in order to keep a roughly constant number of particles per cell. We used a single dust size of $a = 10.24$ cm.
We run the simulations for 50 orbits each and we remove dust particles only inside the Bondi sphere of the planet.
The resulting time evolution of the dust torque is shown in \cifig{fig:resolution}. The lowest-resolution run produces a substantially smaller torque compared with the higher-resolution simulations, indicating that the vicinity of the planet is insufficiently resolved.
In contrast, the two highest-resolution runs yield very similar torque values. The intermediate-resolution setup corresponds to approximately $10$ squared cells per Hill radius, while the highest-resolution run doubles this value together with the number of particles. The close agreement between these two runs indicates that the dust torque has reached numerical convergence at our fiducial resolution. 
\section{Torque contribution convergence}
\begin{figure}
    \centering
    \includegraphics[width=\linewidth]{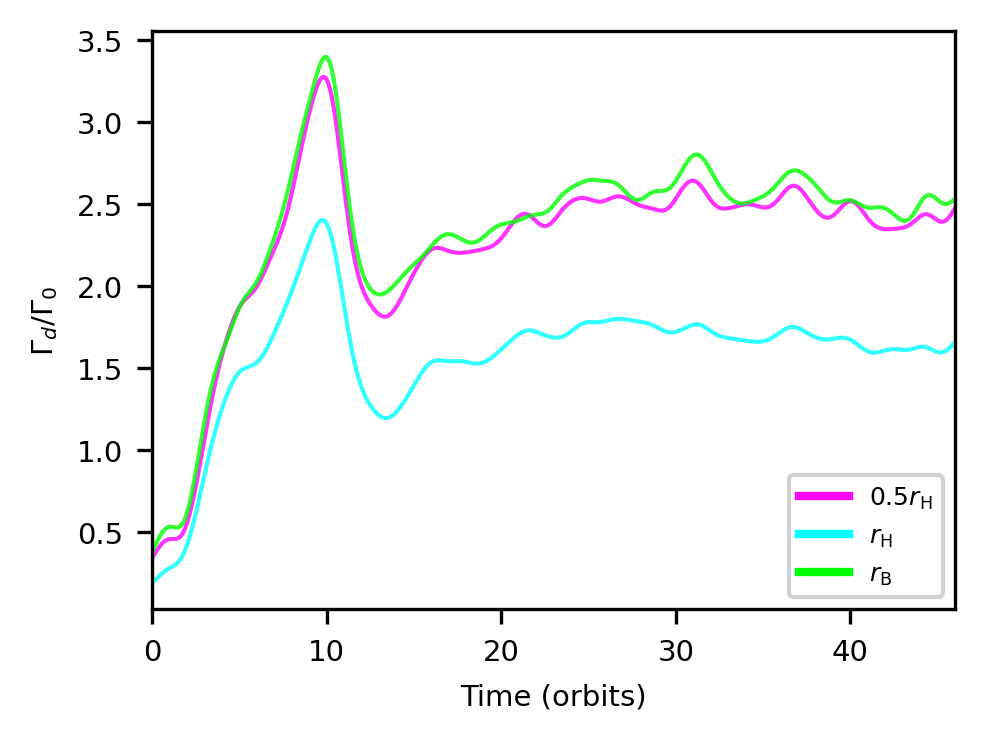}
    \caption{Same as the solid line in \cifig{fig:resolution} but for three different cutoff radii in the torque evaluation.}
    \label{fig:exclusion}
\end{figure}

Previous works exclude part of the material within the Hill sphere of the planet from the torque evaluation \citep{benitez2018, chrenko24}. We tested different cutoff radii to check the convergence of the results. In these tests we do not apply the energy constraint on dust particles, which is considered separately in Appendix \ref{bound}. However, we still remove the particles entering the Bondi sphere, as they are all expected to be accreted \citep{2018MNRAS.479.5136P}. \cifig{fig:exclusion} shows the resulting dust torque for the test run with $N_{\rm{R}} \times N_{\rm{\varphi}} = 351 \times 3675, N = 1 \times 10^6$, and the torque cutoff within $r <[r _{\rm{H}}, 0.5 \, r_{\rm{H}}, r_{B}]$. For M$_{\rm{pl}}$ = 5 M$_{\oplus}$, the Hill radius is $r_{\rm{H}} \simeq 0.0171$ au and the Bondi radius is $r_{\rm{B}} \simeq 0.0059$ au, which means that $r_{\rm{B}} \simeq 1/3 \, r_{\rm{H}}$. As a consequence, neglecting the Bondi sphere or half of the Hill sphere does not change the torque significantly. However, neglecting the full Hill sphere reduces the dust torque by a factor of 1.5. 

\begin{figure}
    \centering
    \includegraphics[width=\linewidth]{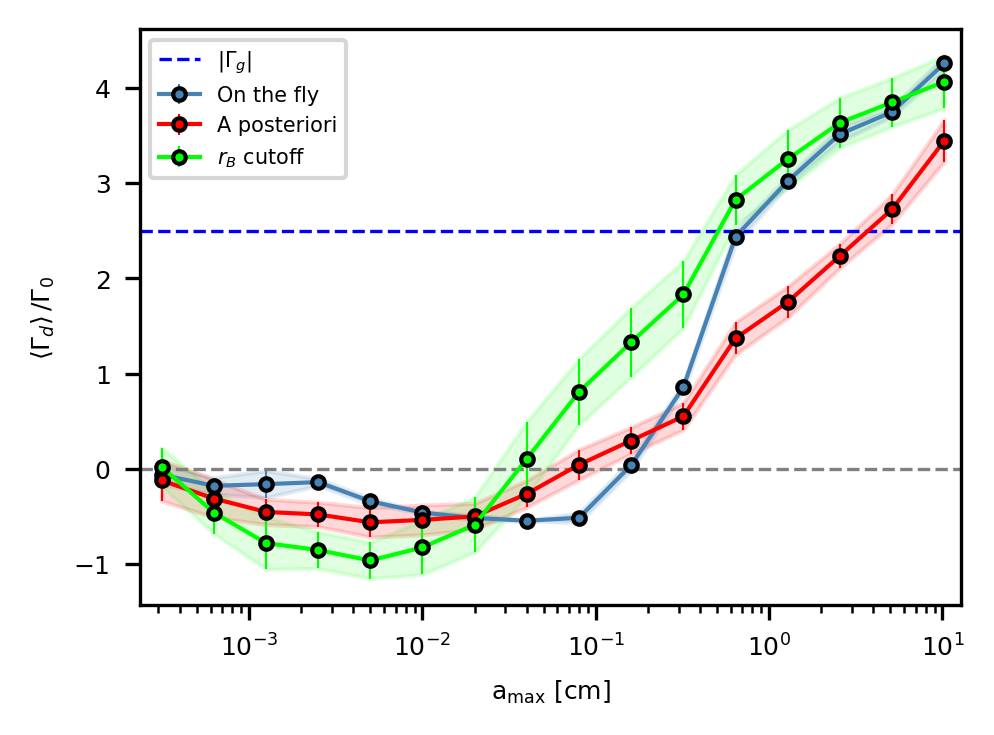}
    \caption{Comparison of the dust torque acting on a $5\,M_{\oplus}$ planet when gravitationally bound particles are removed during the simulation (\emph{on the fly}) or excluded only during the torque evaluation (\emph{a posteriori}). Green line shows the case with only the radial cutoff at $r_{B}$, similar to \cifig{fig:exclusion}. Blue dashed line indicates the absolute value of the gas torque.}
    \label{fig:bound}
\end{figure}

\begin{figure*}[t]
\centering
{\includegraphics[clip]{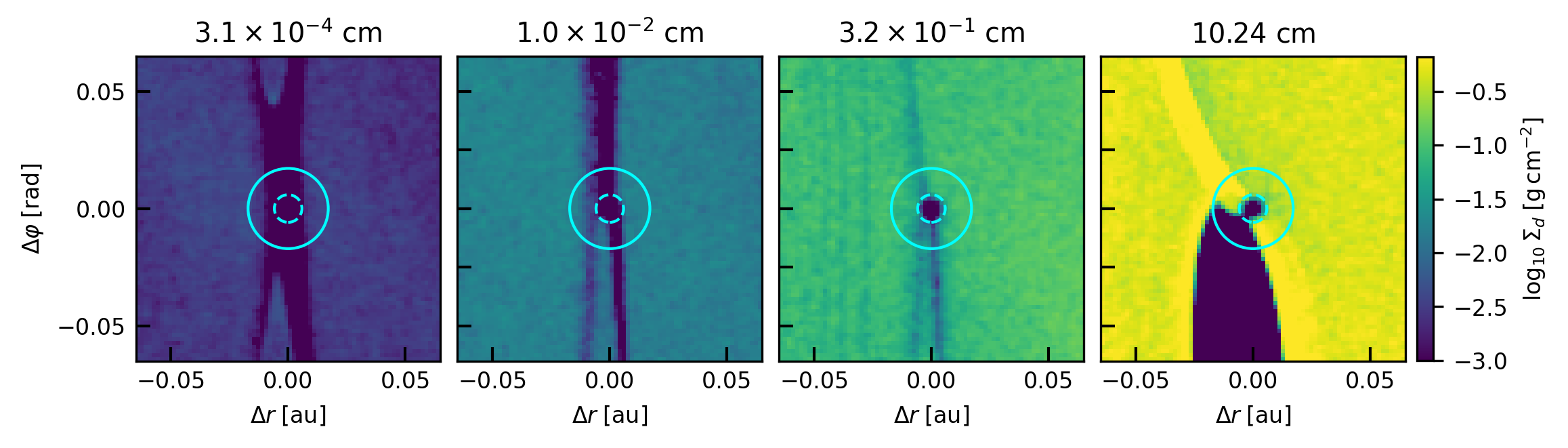}}
\centering
{\includegraphics[clip]{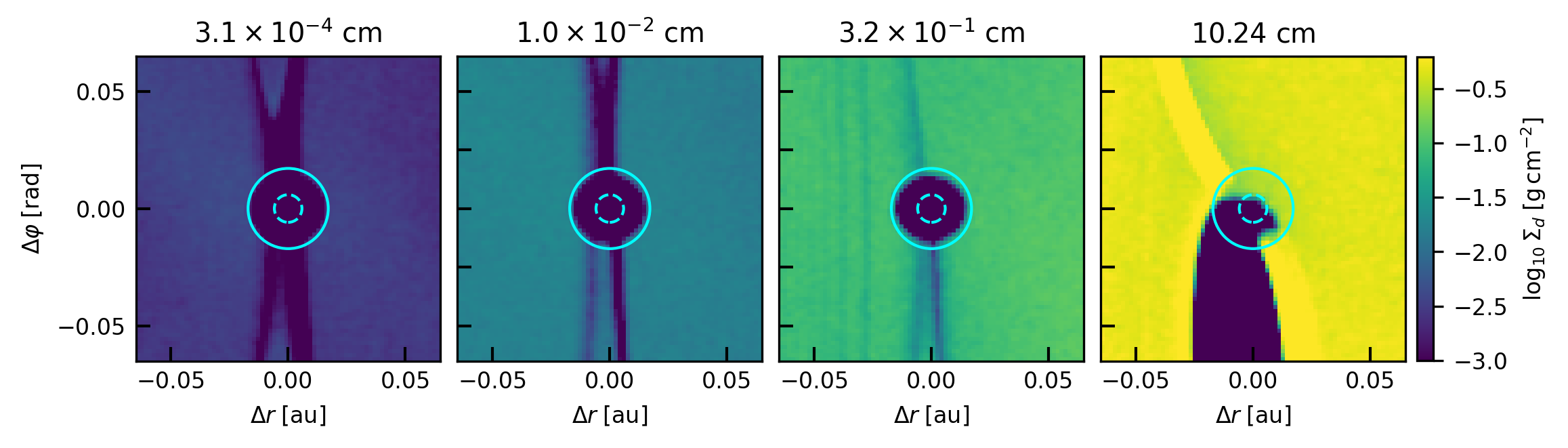}}
    \caption{Top: same as \cifig{fig:dust_dens}, but without the energy-based exclusion criterion. We removed the streamlines to better visualize the dust density. Bottom: same as \cifig{fig:dust_dens}, but we apply the energy-based exclusion criterion a posteriori.}
    \label{fig:bound_density}
\end{figure*}

\subsection{Bound particles} \label{bound}
Rather than introducing an arbitrary cutoff radius, we remove from the simulation the dust particles inside the Hill sphere whenever their kinetic energy is insufficient to escape the planetary potential. 
An alternative approach would be to retain all particles during the simulation and exclude the gravitationally bound ones only \emph{a posteriori} when evaluating the torque. We compare both methods in \cifig{fig:bound}. The two approaches produce qualitatively similar trends, but removing bound particles \emph{on the fly} yields systematically larger positive torques. In particular, the dust torque exceeds the gas torque for $a_{\rm max} \gtrsim 1$ cm.  {We compare the energy-based exclusion criterion with the radial cutoff at $r_{B}$ (green curve in \cifig{fig:bound}) and we find similar results when the gravitationally bound particles are removed during the simulation. In particular, the radial cutoff represents the worst-case scenario in which gravitationally bound particles are contributing to the torque, while the energy-based criterion provides a more conservative estimate, preventing the circumplanetary flow to contaminate the torque measurements. However, this approximation has potential drawbacks due to the dissipative nature of gas drag, which affects the instantaneous energy of the dust particles. In fact, a particle may become temporarily bound to the planet and still escape from its gravitational potential at later times. In this case, the energy-based criterion removes the dust particle and its contribution to the torque from the simulation. This can be seen by comparing \cifig{fig:dust_dens} in the main text to \cifig{fig:bound_density}, where we show the dust density computed using the radial cutoff (top), or applying the energy-based criterion a posteriori (bottom). While the distribution of small and large grains remains similar, we notice significant differences in the 0.32 cm dust size (or equivalently St = 10$^{-2}$), which causes the slight differences in the resulting torque in \cifig{fig:bound}. These artificial density perturbations are undesirable but difficult to avoid with our chosen approach. A more accurate treatment of pebble accretion would require a higher-precision integration scheme capable of reliably identifying irreversible capture by the planet, and is therefore left for future work.}

\end{appendix}

\end{document}